# Synchronous replication of remote storage

*Timur Mirzoev*

*(Department of Information Technology, Georgia Southern University, Statesboro 30460, USA)*

**Abstract:** Storage replication is one of the essential requirements for network environments. While many forms of Network Attached Storage (NAS), Storage Area Networks (SAN) and other forms of network storage exist, there is a need for a reliable synchronous storage replication technique between distant sites (> 1 mile). Such technology allows setting new standards for network failover and failback systems for virtual servers; specifically, addressing the growing need for effective disaster recovery (DR) planning. The purpose of this manuscript is to identify newest technologies such as SAN/iQ and Storage VMotion that allow for remote storage synchronous replication for virtual servers. This study provides an analysis and a comparison of various SANs that create solutions for enterprise's needs. Additionally, the interoperability of these technologies with the industry's leading product VMware ESX Server will be discussed.

## 1. Introduction

With much advancement in the sector of local area networks (LAN), there is still a growing need for efficient storage networks. Data repositories grow every second while each type of storage system may be more or less efficient depending on applications. IDC estimates 2007 revenue for the worldwide storage services market to be $31.7 billion, forecast to grow to $39.9 billion by 2012[1]. The tremendous expansion of storage area networks created new technologies for SAN management and administration. Traditionally, scalability of storage networks has been solved "on-the-fly", i.e. adding more storage components to LANs, which creates gaps in hardware utilization, I/O processing times and manageability of storage networks. Today, SAN administrators more than ever face the challenges of the fast growing rates of specific data repositories that require reliable data replication with failover and failback network systems. Server virtualization has been rapidly changing in the past 5 years, whereas application and network virtualization is not expected to have many changes in the next 10 years[2]. Companies such as LefthandNetworks utilize TCP/IP protocol for extending storage systems over IP and responding to the needs of virtual servers. For example, LefthandNetworks provides an excellent approach called Thin Provisioning. According to IDC, "thin provisioning, also known as oversubscription, allows administrators to maintain a single free space buffer pool to service the data growth requirements for all applications in a shared storage configuration"[3]. EMC and VMware work together on many solutions that address the world of virtualization of storage area networks. In this ever-changing environment, the need for storage replication is even more important. Downtime directly results in financial loss and for many types of businesses it is unacceptable. Recovery Time Objective (RTO) becomes golden for many enterprise systems with a need for high availability. Various techniques of storage replication strive to address concerns for fast RTOs. This manuscript presents the benefits of synchronous replication of remote storage as one of the advanced approaches in storage replication.

## 2. Synchronous vs. asynchronous

Disaster occurrence may be caused by systems failures, human errors or infrastructure failures - only a

Timur Mirzoev, professor, he has over 10 years of experience in information technology, administration and higher education; research fields: server and network storage virtualization, information systems and security, LAN, WLAN networks security.

few to consider for DR planning and each one of them may create a need to recover systems in a short period of time. If all remote storage targets are kept synchronized at all times, RTO can be minimized and systems may achieve 100% availability. Appropriate RTO and Recovery Point Objective (RPO, the point in time to which a systems needs to be recovered) change significantly with the replication approach. For example, RPO is zero for synchronous replication and for asynchronous replication it may be small, depending on the network performance and storage systems availability. According to ITI Inc., large RTO translates to business' high tolerance to a "great loss of data"[4]. Table 1 presents the benefits and drawbacks of using synchronous or asynchronous replication of storage.

Each approach has its benefits and drawbacks. In general, Synchronous replication requires higher investment rates but provides zero recovery time (RTO sometimes is as long as the "restart time"). Distance limitations practically do not exist for asynchronous replication which provides safer data repositories in case of disastrous events.

**Table 1    Synchronous replication vs. asynchronous**

| Replication Type | Distance | Bandwidth | RPO | Availability |
|---|---|---|---|---|
| *Synchronous* | up to 150 miles | High bandwidth, available 100% | Zero RPO (replica is identical at all times) | 100 % |
| *Asynchronous* | Global availability | Average | Small RPO | Depends on RTO |

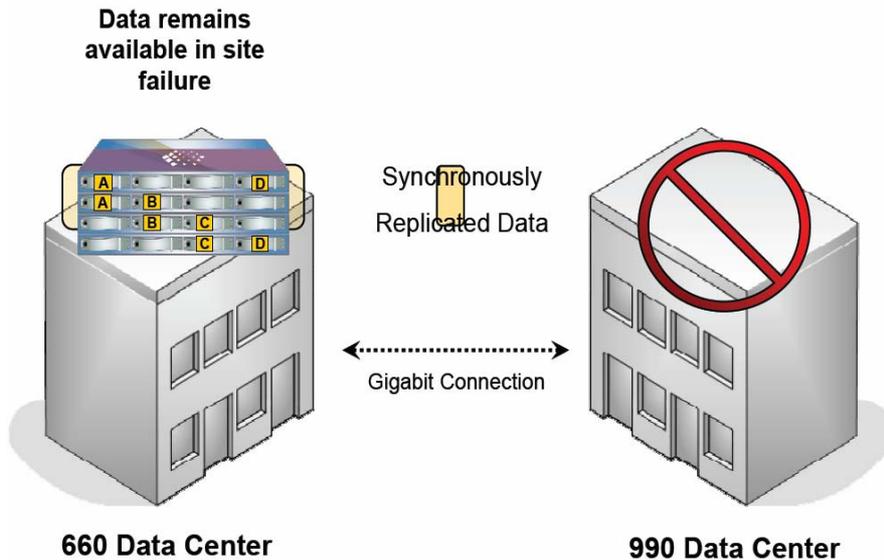

**Fig. 1   Denver Health Campus SAN deployment**

Source: http://inet-gw.maphis.homeip.net/training/VMWare/VMWorld-2007/Sessions/BC/PS_BC26_289343_166-1_FIN_v4.pdf.

SAN technology deployed at Colorado's healthcare institution, Denver Health, could serve as a vivid example of synchronous replication of storage for disaster recovery. Denver Health paired two technologies – virtualization of servers, using VMware's ESX High Availability and LeftHand

SAN/iQ Multi-Site/Campus SAN[5]. According to David Boone, the Operations and Planning Manager of Denver Health, the institution saved about $260,000 (running about 100 servers on 6 ESX machines) and "users don't notice when we're testing"[5]. Denver Health implemented synchronous replication of storage via low latency Gigabit Ethernet connection between two remote sites that are several blocks away with failover and failback systems and with physical separation of storage clusters[6]. Fig. 1 presents a diagram of the storage replication deployment at Denver Health.

For Denver Health, simplicity of DR management, SAN reliability and scalability alone with direct saving on hardware resulted from the integration of server and storage virtualization.

Another example of remote replication is by Mitsubishi Electric Automation that runs SAP ECC6.0 database using SAN/iQ Snap, SAN/iQ Advanced Provisioning and SAN/iQ Remote Copy. Mitsubishi Electric Automation was looking for redundancy, management and performance in its business operations[7]. In the future, according to LefthandNetworks, Mitsubishi is considering adding about 15TB of data for DR, CRM and other applications[7].

## 3. Virtualization of storage

Virtualization concept is not a new idea. It is based on a time-sharing concept originally developed by scientists at Massachusetts Institute of Technology (MIT) in 1961[8]. Server virtualization has rapidly advanced to the level where network storage systems must be available at 100% of time and provide reliability, manageability and scalability. Highly scalable systems save money in a long run but may require higher initial investments. Although many companies claim high scalability, SAN administrators immediately know the limitations of deployed SANs as soon as storage space becomes inadequate. Popularity of VMware's Virtual Infrastructure 3 (VI3) enterprise solution progresses the advancement of virtualization of server to the next level – Distributed Resource Scheduling (DRS), High Availability (HA), Enhanced HA and Distributed Power Management (DMP). These are some of the essential technologies provided by VI3. One of the major components for VI3 is not just ESX machines and Virtual Infrastructure Management; it is the networked storage behind virtualization of servers. The availability of 100% for virtual servers is highly depended on the availability of SAN running in the background of ESX servers. LANs are usually viewed as the "front-end" networks whereas SANs are considered "back-end". Migration of virtual machines "live", without interruption of service is only possible if SAN is available at 100% of the time. The importance of network storage has been stressed many times; however, with technologies such as virtualization of servers, the rules of enterprise networking are constantly changing. Various Fibre Channel (FC) technologies for network storage exist today but the certain limitation such as high cost, scalability and proprietary hardware forced further advancement of Ethernet. IP-based systems provide great scalability and standards. Protocols such as FCIP, iFCP provide vast benefits to enterprise network storage systems but also have certain limitations. FC-based SANs are great when there is no need to extend SAN over a distance. As soon as the need for distance is involved, IP-based data provisioning proves to be more efficient. Perhaps for those reasons, iSCSI evolved a protocol that does not involve utilization of any FC equipment (i.e. all Ethernet-based). iSCSI simply transports SCSI commands over TCP/IP[11]. Ethernet-based systems are cost-effective and highly scalable systems but they are limited by the bandwidth of the Ethernet channel. Today, 10Gb/s standard is no longer the "golden" bandwidth – 40Gb/s and 100Gb/s transfer rates are soon to become new standards[9].

Founded in 1999, LefthandNetworks is the company that takes Ethernet-based iSCSI SANs virtualization to a new standard - SAN/iQ. SAN/iQ technology was developed in 2001 and it is x86-based highly scalable IP SAN[10]. Fig. 2 presents the logical connectivity of SAN/iQ systems. LefthandNetworks calls its IP SAN storage units Virtual Storage Appliance (VSA). VSAs connect various network storage clusters as one while providing excellent SAN administration tools. For example, physical disks could be added or removed from SAN and the SAN/iQ management software simply sees the change and distributes storage resources without any need for administrator's attention[11].

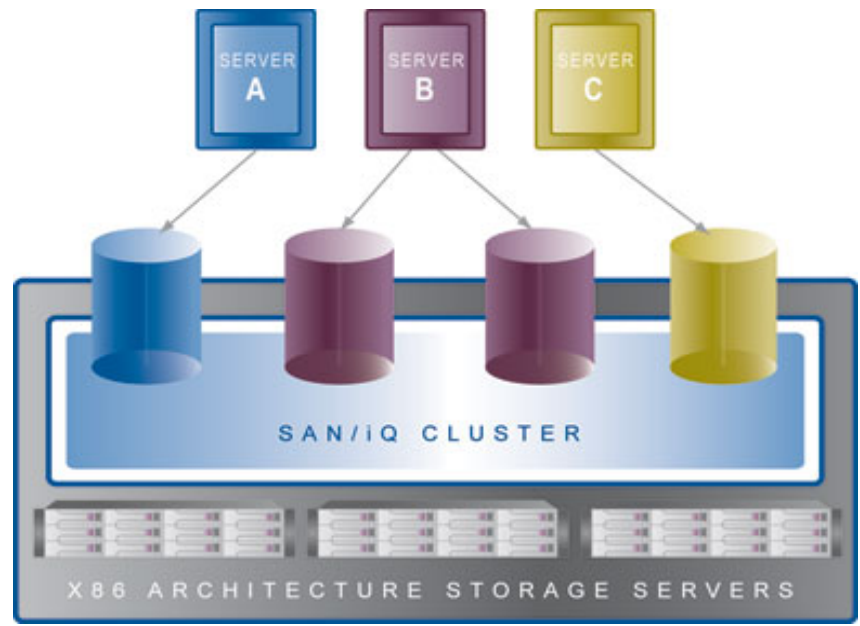

**Fig. 2   SAN/iQ structure**

Source: LefthandNetworks (2007). White paper: Matching Server Virtualization with Advanced Storage Virtualization.

Automatic failover and failback is also handled in a similar fashion - once the Primary Site is down, it automatically fails over a low-latency Ethernet connection to the Recovery Site. After the operations of the Primary site are restored, SAN/iQ acknowledges the response from the Primary site and fails the network storage back from the Recovery Site[11]. The advantage of using SAN/iQ is based in the way network storage is viewed. All storage resources may be viewed as one resource pool that allows for simplification of SAN management, remote replication, high scalability and reliability. According to InfoStor lab test, "SAN/iQ goes as far as entirely handling I/O load - balancing, which is critical for cluster scalability"[12].

The need for reliable SANs provisioning for virtual machines requires storage to be accessible at all times; moreover live migration of Logical Unit Numbers (LUNs) without interruption of services is essential. Fig. 3 illustrates how SAN/iQ technology provisions SAN clusters for VMware ESX servers.

LefthandNetworks is not the only company that does storage virtualization. VMware is working on the Storage VMotion technology that virtualizes storage resources across IP-based networks. Fig. 4 presents the Storage VMotion logical connectivity. The technology is similar to SAN/iQ clustering.

With a successful deployment of virtualized IP SANs, remote replication and disaster recovery become extremely simplified and reliable. However, according to LefthandNetworks, there are certain limitations for SAN/iQ-based synchronous replication deployment: 1) low latency (2ms), 2) high bandwidth, 3) distance up to 20 miles [11].

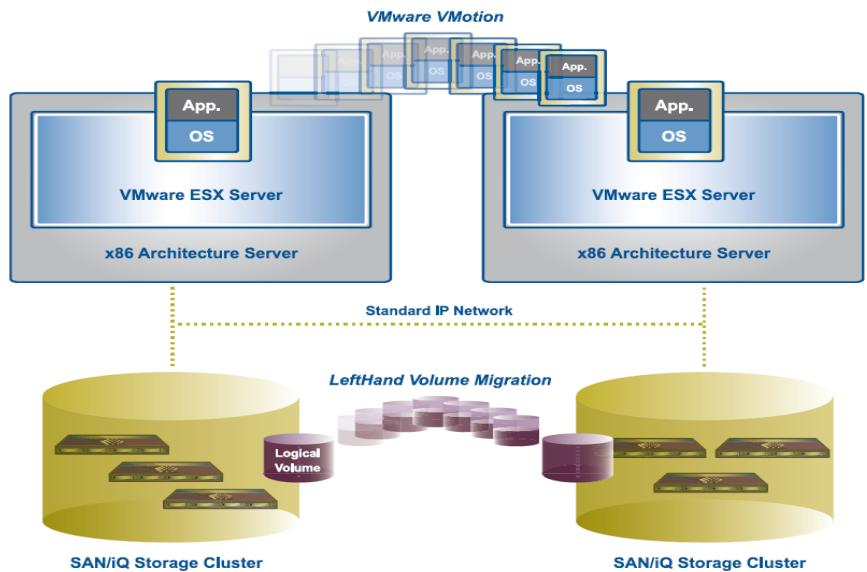

**Fig. 3   SAN/iQ cluster provisioning**
Source: http://www.lefthandnetworks.com/products/index.php.

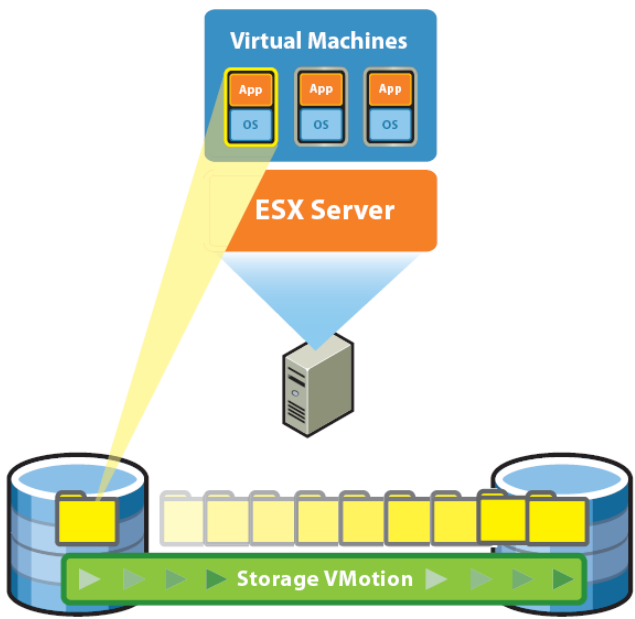

**Fig. 4   Storage Vmotion**
Source: http://www.vmware.com/products/vi/storage_vmotion _overview.html

## 4. Conclusion

Synchronous replication of storage is not a novel technology, but today it has progressed to the level where virtualized SANs increase return-on-investment (ROI) while providing effective solutions for storage replication, disaster recovery and business continuity. The biggest advantages of virtualization technology of storage are the effective use of resources, availability, scalability and manageability. Scalability becomes essential to network storage administrators when systems have to be reconfigured, upgraded or migrated to a different physical location. IP SANs, specifically SAN/iQ, provide highly scalable solutions while SAN management is simple with increase in reliability. Today, server virtualization is essential but it is time to realize the full benefits of storage virtualization technologies such as VSA (SAN/iQ) and storage VMotion which compliment server virtualization with many benefits.

**References:**

[1]   Chandler D. Worldwide Storage Services 2008–2012 Forecast. *Storage and Data Management Services*, IDC, 2008.

[2]   Schreck G. Forrester TechRadar™: Infrastructure Virtualization, *Forrester Research,* 2008.

[3]   Enterprises. *IDC Filing Information*, 2007.

[4]   Highleyman B., Holenstein P., Holenstein B. Replicating Applications for Disaster Recovery. 2003.Available at: http://www.iticsc.com.

[5] Denver Health. *VMware Success Story: Denver Health.* 2007. Available at: http://www.vmware.com/files/pdf /customers/08Q2_ss_vmw_denver_health_english.pdf.

[6] Boone D., Scott B. How Denver Health is Using VMware and LeftHand Networks Multi-Site SAN for Simple, Cost Effective Disaster Recovery. *VMworld 2007 BC26.* 2008. Available at: http://inet-gw.maphis.homeip.net/training /VMWare/VMWorld-2007/Sessions/BC/PS_BC26_2893 43_166-1_FIN_v4.pdf.

[7] LefthandNetworks. Case Study: Mitsubishi Electric Corporation, 2007. Available at: http://www.lefthandnetworks.com/document.aspx?oid=a0 e000000000064AAA.

[8] Corbató F., Daggett M., Daley R. An Experimental Time-Sharing System. Available at: http://larch-www. lcs.mit.edu: 8001/~corbato/sjcc62. 1963.

[9] Waxman J., Yezhkova N., DuBois L. LeftHand Networks —Delivering Open iSCSI SANs for Midmarket and Dornan A. One network to rule them all. *InformationWeek,* May 18, 2008.

[10] LefthandNetworks. White paper: Matching Server Virtualization with Advanced Storage Virtualization. 2007.

[11] LefthandNetworks. Matching Server Virtualization with Advanced Storage Virtualization. Available at: http://www.lefthandnetworks.com/webcasts.aspx.

[12] Fegreus J. SANitize DAS to enhance virtual infrastructure. *InfoStor Lab Review,* 2008.Available at: http://www. infostor.com/article_display.content.global.en-us.articles. infostor.top-news.lab-review-sanitize-das-to-enhance-virt ual-infrastructure.1.html.

[13] Microsoft Storage. Storage Glossary: Basic Storage Terms, 2005. Available at: http://www.microsoft.com /windowsserversystem/storage/storgloss.mspx.

(Edited by Alex, Amy)

[6] Antonioli F. n-Track Studio. Retrieved 3rd of October, 2007. Available at: http://www.fasoft.com/. 2007.

[7] NoteWorthy Software, Inc. NoteWorthy Composer Retrieved 3rd of October, 2007. Available at: http://www.noteworthysoftware.com/. 2007.

[8] GoldWave, Inc. GoldWave. Retrieved 3rd of October, 2007. Available at: http://www.goldwave.com/. 2007.

[9] Song of the year. Songwriting Songwriter Contest Supporting VH1. Retrieved 3rd of October, 2007. Available at: http://songoftheyear.com/. 2004.

[10] The UK songwriting contest. UK national songwriting competition. Retrieved 3rd of October, 2007 Available at: http://www.songwritingcontest.co.uk/. 2004.

[11] *TIME Magazine.* 50 Best Websites. Retrieved 3rd of October, 2007. Available at: http://www.time.com/time /techtime/200306/intro.html. 2003.

[12] Garageband. GarageBand Discovering the best independent music. Retrieved 3rd of October, 2007. Available at: http://www.garageband.com/. 2004.

[13] Toronto Experimental Artists. Various Artists compilation CD Vol 2. Retrieved 3rd of October, 2007. Available at: http://www.teasouth.com/, 2004.

[14] Premis Industries Carpe Diem 2. Retrieved 3rd of October, 2007. Available at: http://www.premisind.com/products /carpediem2.html. 2005.

[15] Dbfoto. Liquid Adrenaline. Retrieved 3rd of October, 2007. Available at: http://www.dbfoto.com. 2004.

[16] SavageDogFilms. Rip Cage. Retrieved 3rd of October, 2007. Available at: http://www.savagedogfilms.com /ripcage. 2005.

[17] CNET. Download.com. Retrieved 3rd of October, 2007. Available at: http://music.download.com, 2007.

[18] International Marketing Group. IMG Music Promotions. Retrieved 3rd of October, 2007. Available at: http://clix.to/img, 2004.

[19] Caragan Music Agency. Caragan Music Agency. Retrieved 3rd of October, 2007. Available at: http://wwwcaragan.co.uk/. 2005.

[20] The Unleaded. The Unleaded Band Official Website. Retrieved 3rd of October, 2007. Available at: http://www.TheUnleaded.com. 2007.

(Edited by Flyer, Amy)